\documentclass[3p,times,twocolumn]{elsarticle}
\usepackage{amssymb}
\usepackage{amsbsy}
\usepackage{graphics}
\usepackage{psfrag}
\usepackage{epsfig}
\usepackage{multirow,mathtools}
\usepackage{bm}
\usepackage{color}
\usepackage{soul}
\usepackage{amsmath, bbm}
\usepackage[figuresright]{rotating}
\usepackage{capt-of}

\newcommand{\La}{{\Lambda}}
\newcommand{\Si}{{\Sigma}}

\newcommand{\be}{\begin{eqnarray}}
\newcommand{\ee}{\end{eqnarray}}

\begin{document}

\begin{frontmatter}
\title{Exploring the $\Sigma^+ p$ interaction by measurements of the correlation function}
 \author[J]{Johann Haidenbauer}
 \author[B,J,C]{Ulf-G. Mei{\ss}ner}

 \address[J]{Institute for Advanced Simulation, Institut f{\"u}r Kernphysik and
 J\"ulich Center for Hadron Physics, Forschungszentrum J{\"u}lich, D-52425 J{\"u}lich, Germany}
 \address[B]{Helmholtz Institut f\"ur Strahlen- und Kernphysik and Bethe Center
  for Theoretical Physics, Universit\"at Bonn, D-53115 Bonn, Germany}
 \address[C]{Tbilisi State  University,  0186 Tbilisi, Georgia}

\begin{abstract}
The spin-dependent components of the fundamental hyperon-nucleon interactions are largely unknown.
We show that under reasonable assumptions, a measurement of the $\Sigma^+ p$ correlation function
in high-energy proton-proton or heavy-ion collisions combined with cross section data allows to separate 
the singlet and the triplet $S$-wave contributions.
\end{abstract}

%\pacs{13.60.Rj,14.40.Lq,25.43.+t}

\end{frontmatter}

%%%%%%%%%%%%%%%%%%%%%%%%%%%%%%%%%%%%%%%%%%%%%%%%%%%%%%%%%%%%%%%

{\bf Introduction:}~There has been a renewal of interest in hyperon-nucleon 
interactions recently, triggered on the one side by the study of hypernuclei at 
facilities world-wide \cite{J-PARC:2019,JLAB:2018,PANDA:2020}
and on the other side driven by studies of the equation of state
of neutron matter. The latter is  especially of  relevance for understanding
the properties of neutron stars, and in particular for the
generation of gravitational waves in mergers of such compact objects, see, e.g.,
\cite{Annala:2017llu} and the recent reviews \cite{Chatterjee:2015,Tolos:2020aln}. 
In fact, the interactions of hyperons
($Y = \La, \Si, \Xi$) with  nucleons have been studied for many decades.
However, so far only  the bulk properties are experimentally established,
and first of all those for the $\La N$ system. More detailed information, specifically 
about the spin dependence of the $YN$ forces, is completely lacking. 
The latter is a consequence of the short-lived character of the hyperons 
so that no proper beams can
be prepared in order to perform standard scattering experiments. 
An essential additional source of information on the $YN$ force 
is provided by studying their interaction in reactions where the 
two baryons emerge with low relative momentum in the final state  
\cite{Tan:1973,Hinterberger:2004,Budzanowski:2010ib,TOF:2012,Munzer:2018,Wilkin:2016}. 
In such a case, in principle, even the scattering lengths 
can be extracted. Indeed, with that aim in mind the $\La p$ final-state 
interaction in the reaction $pp\to K^+\Lambda p$
has been measured at the COSY accelerator in J\"ulich  
\cite{TOF:2013,Hauenstein:2017}. A determination of the
spin dependence, i.e. the separation of the spin-triplet and/or 
singlet amplitude, is possible in that reaction, but requires 
a single- or even double-polarization experiment 
\cite{Gasparyan:2004,Gasparyan:2005}.
Unfortunately, efforts to determine the strength of the spin-triplet 
$\La p$ interaction by the TOF collaboration \cite{Hauenstein:2017} 
suffered from low statistics and, at the end, did not provide the
anticipated conclusive results.

More recently, it was realized that information on the $YN$ scattering lengths 
can also be obtained from studying the corresponding two-body momentum correlation 
function as measured in heavy-ion collisions or high-energetic $pp$ 
collisions \cite{Shapoval:2015,Cho:2017}.
With respect to $\La p$ there are data from the STAR 
Collaboration \cite{Adams:2006}
from a measurement in Au+Au collisions at $\sqrt{s}=200$~GeV
and by the ALICE collaboration 
in $pp$ collisions at $\sqrt{s}= 7$~TeV \cite{Acharya:2019}
and $\sqrt{s}= 13$~TeV \cite{Acharya:2021}, respectively. 
However, also here no detailed information on the spin dependence 
can be deduced. The singlet- and triplet contributions simply
add up with the same statistical weight as in free scattering, 
as commonly assumed \cite{Shapoval:2015,Acharya:2019}.

Here, we point out that a separation of the spin 
states is feasible from measurements of the correlation function - 
in combination with cross section data - without any spin-dependent 
experiment.
It is possible in specific cases, namely when the interaction in one 
of the spin states is attractive while that in the other one is
repulsive. As will be discussed in more detail below, in such a
situation the contributions of the spin states to the correlation
function will partially cancel, because they depend on the sign of
the scattering amplitude, whereas they always add up in case of
the reaction cross section. This qualitatively different interplay
allows one to disentangle the spin contributions,
provided one has data on the correlation function and on the
elastic cross section. 

We exemplify this idea for the $\Si^+ p$ system, where cross section data at 
low energies have been available for a long time \cite{Eisele:1971}.
This channel has isospin $I=3/2$ so that there is no coupling to
the $\La N$ system and the reaction is purely elastic, i.e.
complications from open channels \cite{Haidenbauer:2018,Kamiya:2021} are absent. 
At low energies, the observables are dominated by the (spin singlet) $^1S_0$ 
and (spin triplet) $^3S_1$ partial waves. 
There is strong evidence that the interaction in the $^3S_1$ partial 
wave is repulsive. First indications for its repulsive nature came from the 
analysis of level shifts and widths of $\Sigma^-$ atoms and from measurements 
of $(\pi^-,K^+)$ inclusive spectra related to $\Sigma^-$-formation in heavy 
nuclei \cite{Gal:2016}.
As discussed in detail in Ref.~\cite{Haidenbauer:2015}, a repulsive 
$\Sigma$ potential in the medium can be only achieved when the interaction 
in the $^3S_1$ partial wave of the $\Si^+ p$ channel is repulsive.
More recently, strong and more direct evidence for a repulsion 
has been provided by lattice QCD calculations \cite{Beane:2012,Nemura:2018}.
The first study was performed for unphysical quark masses, 
corresponding to $M_\pi \simeq 389$~MeV, however, the second is for 
almost physical quark masses ($M_\pi \simeq 146$~MeV). Both studies
support a strongly attractive $^1S_0$ interaction and a weakly repulsive
$^3S_1$. 
Modern $YN$ potentials derived within chiral effective field theory (EFT) 
\cite{Polinder:2006,Haidenbauer:2013,Haidenbauer:2019} 
produce a repulsive $^3S_1$ interaction,
but also some $YN$ interaction models based on the meson-exchange approach
\cite{Haidenbauer:2005,Nagels:2019} or on the constituent-quark model
\cite{Fujiwara:2006}. Note that the strongly attractive 
interaction in the $^1S_0$ channel is a consequence of the underlying 
approximate SU(3) flavor symmetry which closely links that state with the 
likewise strongly attractive $^1S_0$ partial wave in the nucleon-nucleon 
sector \cite{Haidenbauer:2015X}. 

%%%%%%%%%%%%%%%%%%%%%%%%%%%%%%%%%%%%%%%%%%%%%%%%%%%%%%%%%%%%%%%%%%
{\bf Formalism:}~The formalism for calculating the two-particle correlation function from a 
two-body interaction has been described in detail in various publications.
Thus, in the following we provide only a summary of the employed formulae, where
we follow very closely the paper of Ohnishi et al.~\cite{Ohnishi:2016}, 
see also \cite{Cho:2017,Haidenbauer:2018,Haidenbauer:2020}. 
The two-particle momentum correlation function $C(k)$ for two non-identical 
particles with interaction in a single $S$-wave state is given by 
\begin{align}
C(k) = 1 + \int_0^\infty 4\pi r^2 dr\,S_{12}(r) 
\left[
\left|\psi(k,r)\right|^2
-\left|j_0(kr)\right|^2
\right] ,
\label{Eq:LL1}
\end{align}
where $k$ is the center-of-mass momentum of the two-body system. 
$S_{12}$ is the so-called source function \cite{Cho:2017} for which 
we adopt the usual static approximation and represent the source by 
a spherically symmetric Gaussian distribution,
$S_{12}(\bold{r})=\exp(-r^2/4R^2)/(2\sqrt{\pi}R)^3$, 
so that it depends only on a single parameter, the source radius $R$. 
$\psi(k,r)$ is the scattering wave function that can be obtained by 
solving the Schr\"odinger or Lippmann-Schwinger equation for a given 
potential, and $j_0(kr)$ is the spherical Bessel function for $l=0$. 
The wave function is normalized asymptotically to \cite{Ohnishi:2016}
\begin{equation}
\psi(kr) \to  S^{-1} \left[ \frac{\sin(kr)}{kr} + f(k) \frac{\exp(i kr)}{r} \right] 
\label{Eq:LL1a}
\end{equation}
where $f(k)$ is the scattering amplitude which is related to the $S$ matrix 
by $f(k) = (S-1) / 2 i k$. For small momenta $k$, the amplitude can be
written in terms of the effective range expansion, i.e.
$f(k)\approx 1/(-1/a_0 + r_0k^2/2 - i k)$, with $a_0$ and $r_0$ the 
scattering length and the effective range, respectively. 

For illustrating how the sign of the scattering amplitude (scattering length)
enters, let us take a look at the Lednicky and Lyuboshitz (LL) 
model \cite{LL1} where the actual wave function in Eq.~(\ref{Eq:LL1}) is 
simply replaced by the asymptotic form Eq.~(\ref{Eq:LL1a}). Then the correlation 
function can be written completely in terms of the scattering amplitude,
\begin{eqnarray}
&\int_0^\infty 4\pi r^2 dr\,S_{12}(r) 
\left[
\left|\psi(k,r)\right|^2
-\left|j_0(kr)\right|^2
\right] 
\approx \nonumber \\
&\frac{|f(k)|^2}{2R^2} F(r_0) 
+\frac{2\text{Re}f(k)}{\sqrt{\pi}R}\,F_1(x)
-\frac{\text{Im}f(k)}{R}\,F_2(x) \ , 
\label{Eq:LL2}
\end{eqnarray}
where $F_1(x)=\int_0^x dt\, e^{t^2-x^2}/x$ and $F_2(x)=(1-e^{-x^2})/x$,
with $x=2kR$.
The factor $F(r_0) = 1-r_0/(2\sqrt{\pi}R)$ is a correction that accounts for the deviation 
of the true wave function from the asymptotic form \cite{Ohnishi:2016}.
Obviously, there is a contribution proportional to ${\rm Re}f(k)$ so that the correlation
function is sensitive to its sign, i.e. to whether the interaction is attractive or 
repulsive. The term results from the interference between the incoming and outgoing 
waves when the wave function is squared, see Eq.~(\ref{Eq:LL1a}). 
% XX  
Indeed, in general this term dominates the correlation function for small momenta, 
see, e.g., Ref.~\cite{Cho:2017} for a more detailed discussion. 
Thus, for a purely attractive interaction ($a_0 < 0$) 
the correlation function is enhanced, i.e. $C(k) \geq 1$, whereas for a purely repulsive 
interaction ($a_0 > 0$) the correlation function is suppressed, i.e. $C(k) \leq 1$. 

As mentioned, at low energies the $\Si^+ p$ system will be in the spin singlet ($s$) 
and spin triplet ($t$) $S$-wave states.  Then the cross section is given by
$\sigma \sim (|f_s(k)|^2 + 3\,|f_t(k)|^2)/4$, considering the appropriate spin weights.  
Also for the correlation function (\ref{Eq:LL1}) an averaging over the spin has to 
be performed. Following the usual assumption that the weight is the same as for 
free scattering \cite{Shapoval:2015,Cho:2017} leads to 
 
\begin{align}
C(k) = \frac{1}{4} C_{s}(k) + \frac{3}{4} C_{t}(k) \ . 
\label{Eq:LL3}
\end{align}
 
\begin{table*}
\caption{$\Si^+ p$ Coulomb modified scattering lengths 
and effective range parameters (in fm) in the $^1S_0$ ($s$) and
$^3S_1$ ($t$) partial waves. 
    }
    \label{tab:acrc}
    \centering
    \renewcommand{\arraystretch}{1.4}
\vskip 0.3cm 
\begin{tabular}{|c|cccc|}
\hline
interaction & $a^c_s$ & $r^c_s$ & $a^c_t$ & $r^c_t$ \\    
\hline
NLO13(600) \cite{Haidenbauer:2013}    & $-3.56$ & $3.54$ & $0.49$ & $-5.08$ \\
NLO19(600) \cite{Haidenbauer:2019}    & $-3.62$ & $3.50$ & $0.47$ & $-5.77$ \\
J\"ulich '04 \cite{Haidenbauer:2005}  & $-3.60$ & $3.24$ & $0.31$ & $-12.2$ \\
ESC16 \cite{Nagels:2019}              & $-4.30$ & $3.25$ & $0.57$ & $-3.11$ \\
Nagels '73 \cite{Nagels:1973}         & $-2.42\pm 0.30$ & $3.41\pm 0.30$ & $0.71$ & $-0.78$ \\
fss2 \cite{Fujiwara:2006}             & $-2.28$ & $4.68$ & $0.83$& $-1.52$ \\
NLO(sim)                              & $-2.39$ & $4.61$ & $0.80$& $-1.25$ \\
\hline
\end{tabular}
\renewcommand{\arraystretch}{1.0}
\end{table*}

Considering the discussion above, $C_s$ and $C_t$ are expected to have opposite signs 
for the $\Si^+ p$ system so that there will be a partial cancellation 
in the evaluation of the actual correlation function $C(k)$.
Therefore, experimental information on the correlation function, together with
data on the cross section, allow one to disentangle the $^1S_0$ and $^3S_1$ 
contributions, and possibly even to determine their scattering lengths, without
performing any spin-dependent experiments. 

Note that in case of the $\Si^+p$ system there is a caveat, namely
the presence of the Coulomb force. Due to the repulsive Coulomb interaction
between the $\Si^+$ and the proton eventually, for momenta $k\to 0$, 
$C_t(k)$ as well as $C_s(k)$ have to show the characteristics of a repulsive interaction.
It will be interesting to see in how far the Coulomb interaction affects the overall results.
In the presence of the Coulomb interaction modifications of 
Eqs.~(\ref{Eq:LL1}) and (\ref{Eq:LL1a}) are required, see Ref.~\cite{Haidenbauer:2020}.
Specifically, the wave functions $\psi (k,r)$ and $j_0 (kr)$ have to be replaced by the 
corresponding solutions including the Coulomb interaction, 
$\psi^{SC} (k,r)$ and $F_0 (kr)/(kr)$, where $F_0(kr)$ is the regular Coulomb wave 
function for $l=0$. 
 
\begin{figure*}
\begin{center}
\includegraphics[height=090mm]{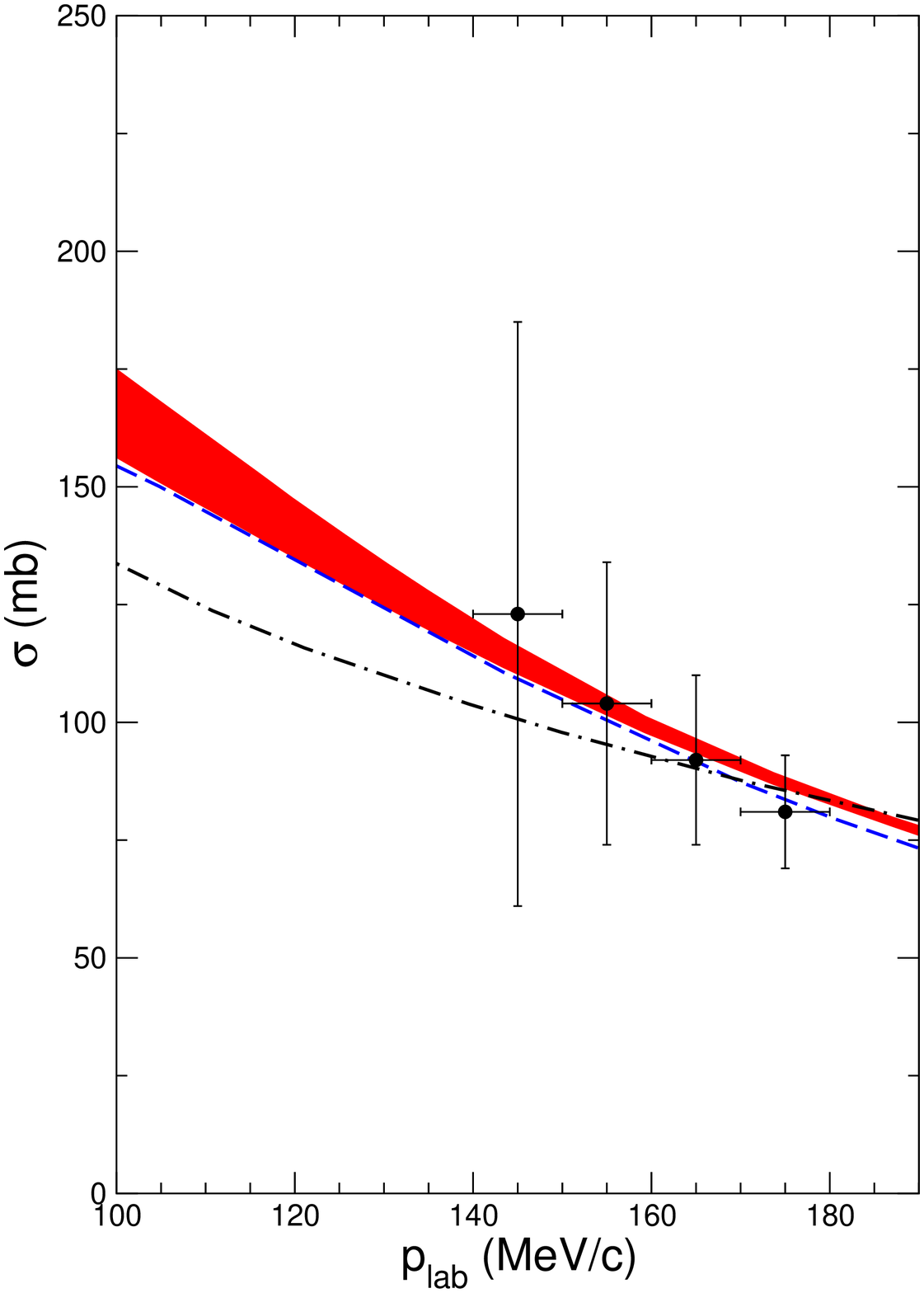}
\includegraphics[height=090mm]{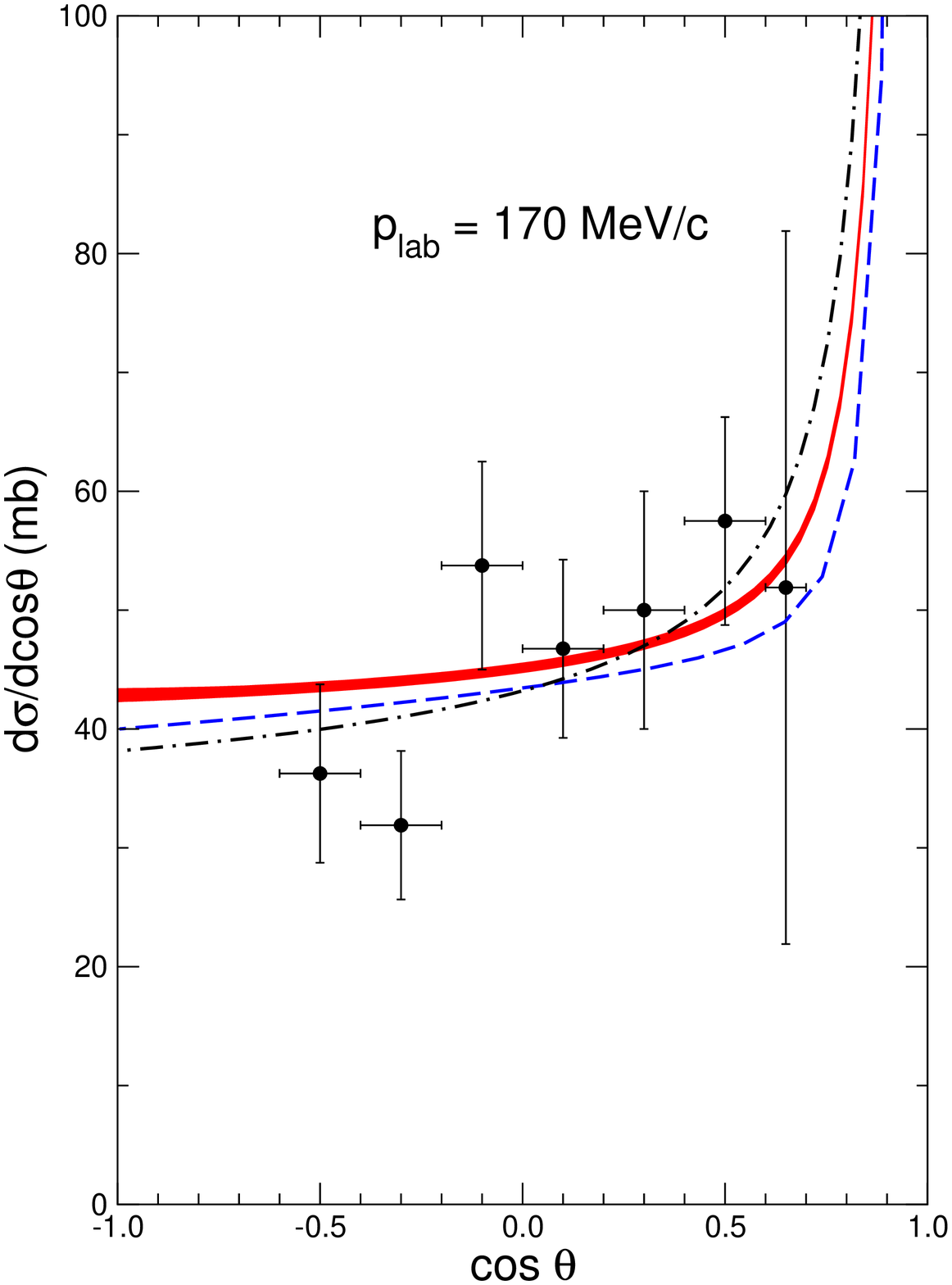}
\vskip -0.5cm 
\caption{Total and differential $\Si^+ p$  cross section.  
The bands represent results from the chiral EFT potential
NLO19 (red-dark) \cite{Haidenbauer:2019}, while the dashed lines are 
those of the J\"ulich~'04 meson-exchange potential \cite{Haidenbauer:2005}. 
The dash-dotted line corresponds to an interaction that simulates  
effective-range parameters as suggested in Refs.~\cite{Nagels:1973,Fujiwara:2006}, 
see Table~\ref{tab:acrc}. The data are from Ref.~\cite{Eisele:1971}. 
}
\label{fig:Sp}
\end{center}
\end{figure*}

% XX
Finally, let us emphasize that the results presented below are all based on 
Eq.~(\ref{Eq:LL1}), i.e. from calculations with the full scattering wave functions.  
When the scattering length is small and the effective range large, 
as is the case for the $\Si^+ p$ $^3S_1$ partial wave, see Table~\ref{tab:acrc}, 
the LL approximation (\ref{Eq:LL2}) is not reliable \cite{Cho:2017,LL1}.  

%%%%%%%%%%%%%%%%%%%%%%%%%%%%%%%%%%%%%%%%%%%%%%%%%%%%%%%%%%%%%%%
{\bf Results:}~For $\Si^+p$ scattering at low energies cross sections but also
angular distributions have been available for a long time \cite{Eisele:1971}. 
The data are shown in Fig.~\ref{fig:Sp}, together with results from $YN$ interactions
derived within chiral  effective field theory (EFT) at next-to-leading order (NLO) 
(specifically, we use the NLO19 version) \cite{Haidenbauer:2019} (red bands)
or based on the traditional meson-exchange picture \cite{Haidenbauer:2005}  
(dashed line). One can see that the data are fairly well reproduced
by these potentials, which actually aimed at a consistent description of 
all low-energy $\La N$ and $\Si N$ scattering data under the assumption
of an (approximate) SU(3) flavor symmetry.    
A more specific analyis of the $\Si^+ p$ data has been presented by Nagels 
et al.~\cite{Nagels:1973}. 
A dedicated phase shift analysis at $p_{lab} = 170$~MeV/c has been attempted 
in \cite{Nagata:2002}.

Selected results for the $\Si^+ p$ scattering lengths from the 
literature are summarized in Table~\ref{tab:acrc}. One can see that 
there are several predictions with a $^1S_0$ scattering length 
of $a_s\approx -4$~fm so that the $^1S_0$ contribution alone practically 
saturates the measured cross section~\cite{Haidenbauer:2015X}. 
Then the $^3S_1$ contribution has to be very small and 
the scattering length amounts to only $a_t \approx 0.5$~fm. 
On the other hand, in the analysis in \cite{Nagels:1973} the singlet scattering 
length is significantly smaller and, accordingly, the one in the triplet state 
noticeably larger. The latter situation is also predicted by a $YN$ potential 
that has been derived within the constituent-quark model \cite{Fujiwara:2006}. 

In order to quantify the impact of such differences on $\Si^+ p$
observables and specifically on the correlation function we constructed
a $\Si N$ interaction within chiral EFT that simulates the latter situation. 
To be concrete, we adopted the
NLO19 potential with cutoff $\La = 600$~MeV from \cite{Haidenbauer:2019} 
and re-adjusted the low-energy constants (LECs) corresponding to the
SU(3) $\{27\}$ and $\{10\}$ irreps \cite{Haidenbauer:2013} to get
values in the range suggested by Refs.~\cite{Fujiwara:2006,Nagels:1973},
see the entry denoted as NLO(sim) in Table~\ref{tab:acrc}. No attempt was made to 
reproduce one or the other result of these works exactly. However, we made sure that 
the total $\chi^2$ (based on the cross section) for NLO(sim) and for NLO19(600) are 
identical (i.e. $\chi^2=0.4$ in both cases). The pertinent results for 
$\Si^+p$ are indicated by the dash-dotted lines in Fig.~\ref{fig:Sp}. 
One can see that the total cross section for the smaller $a_s$ (larger $a_t$) scenario
exhibits a slighly weaker momentum dependence, see also \cite{Nagels:1973,Fujiwara:2006},
while the angular dependence is slighly more pronounced. 
This result suggests that, in principle, a very accurate
measurement of the $\Si^+p$ cross section itself could already allow to 
disentangle the singlet- and triplet contributions. However, as argued above, 
since the contributions add up there is less sensitivity. Furthermore, and more 
decisive, measuring $\Si^+p$ elastic scattering so close to threshold is 
practically impossible. 

\begin{figure}
\begin{center}
\includegraphics[height=85mm,angle=-90]{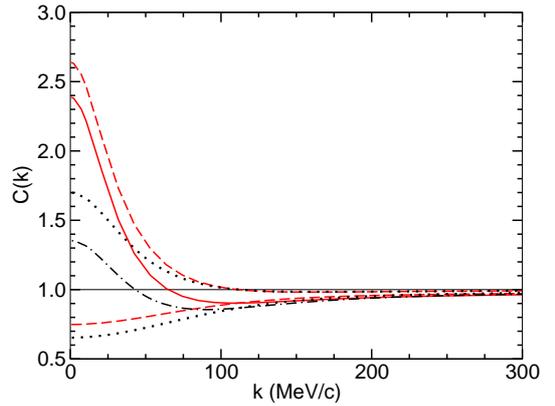}
\caption{Correlation function for $\Si^+ p$ without Coulomb interaction 
predicted by NLO19(600) (solid line) and NLO(sim) (dash-dotted line).  
Contributions from the $^1S_0$ (upper curves) and $^3S_1$ (lower curves) 
alone are indicated by the  dashed (NLO19(600)) and dotted (NLO(sim)) curves. 
The used source radius is $R=1.2$~fm.
}
\label{fig:spp}
\end{center}
\end{figure}

\begin{figure*}
\begin{center}
\includegraphics[height=80mm,angle=-90]{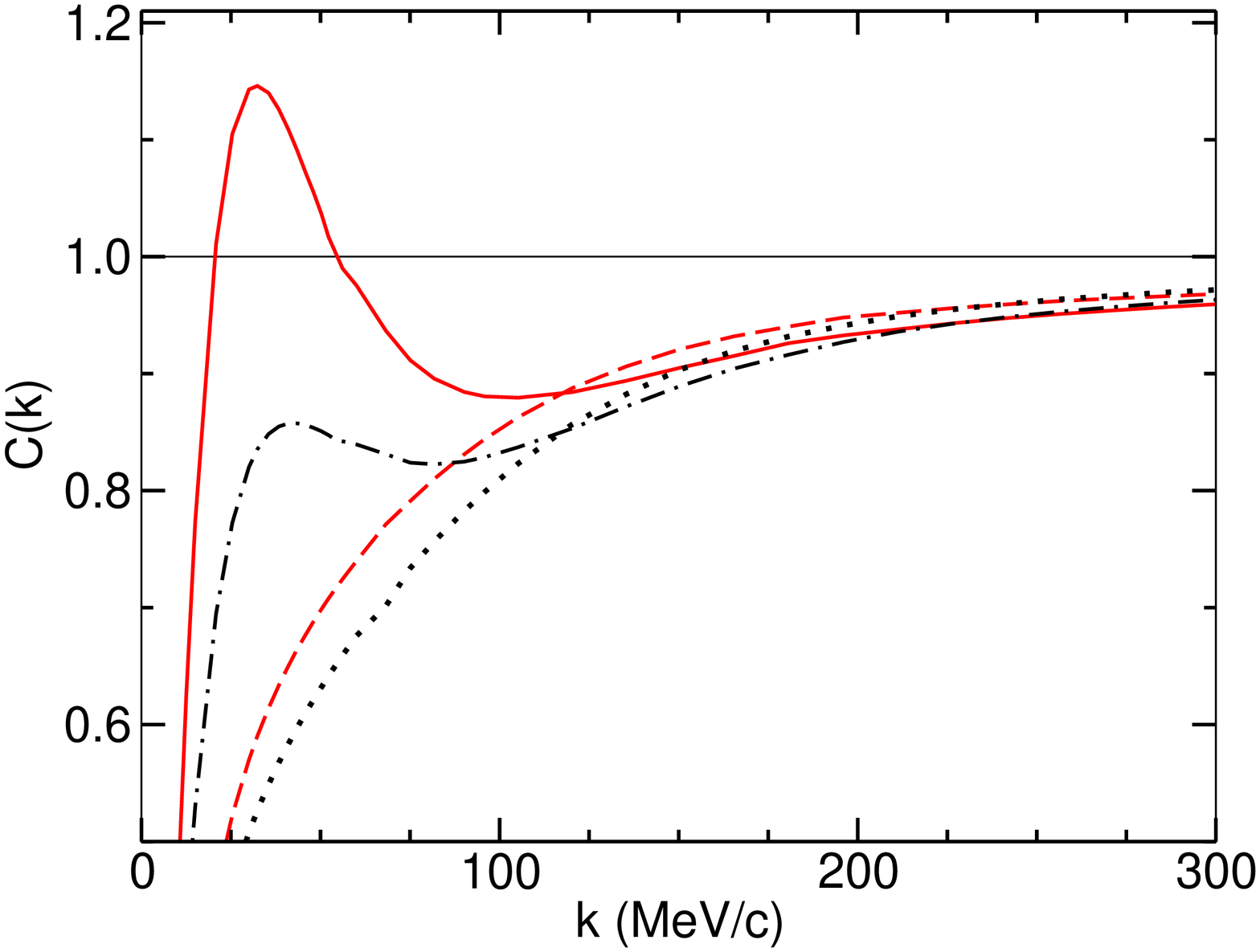}
\includegraphics[height=80mm,angle=-90]{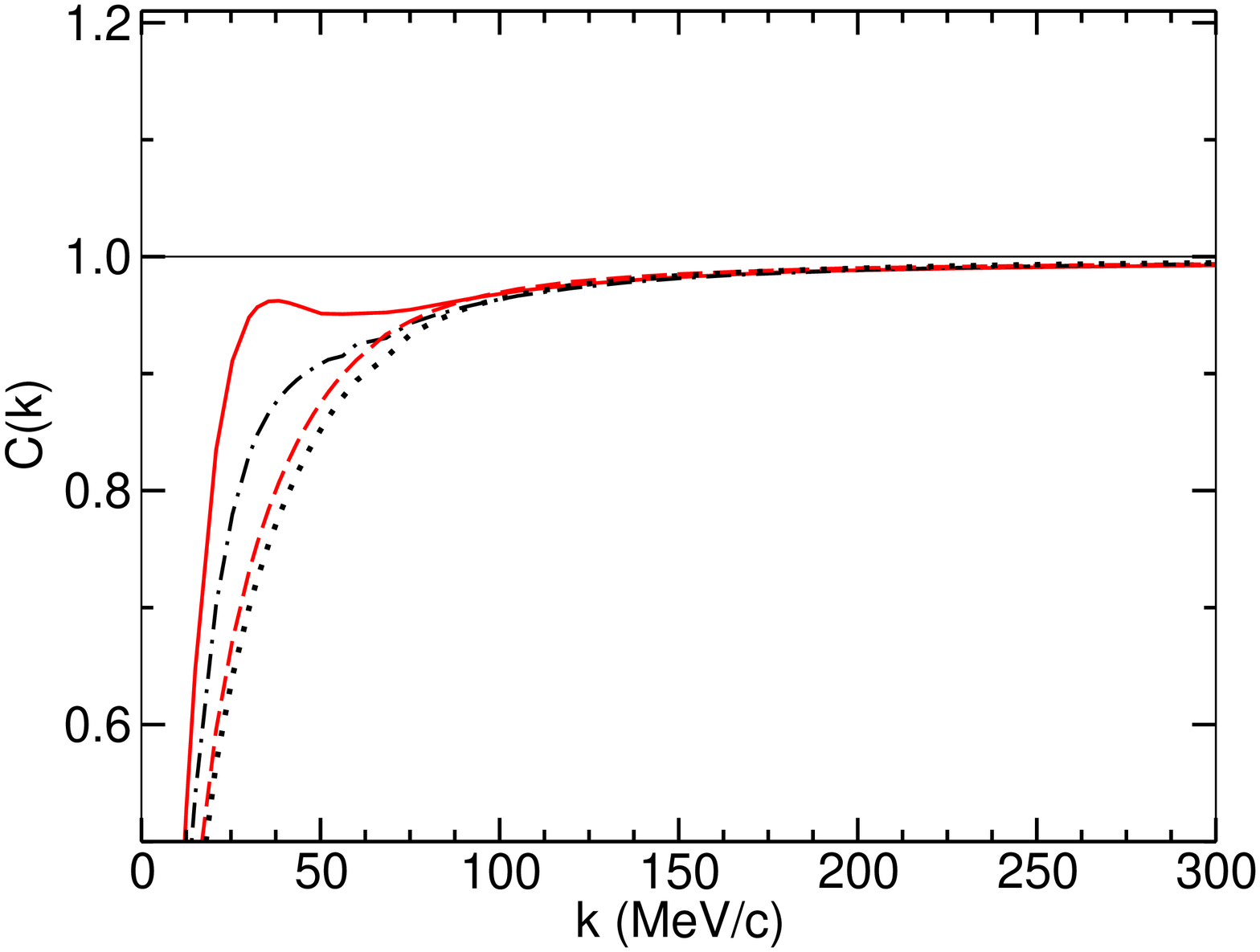}
\caption{Correlation function for $\Si^+ p$ with Coulomb interaction  
predicted by NLO19(600) (solid lines) and NLO(sim) (dash-dotted lines).  
Results for the $^3S_1$ alone are indicated by the dashed (NLO(600) 
and dotted (NLO(sim)) curves. 
Left panel is for the source radius $R=1.2$~fm, right panel for $R=2.5$~fm. 
}
\label{fig:spp1}
\end{center}
\end{figure*}

Let us now look at the predictions for the correlation function. 
A calculation without inclusion of the Coulomb interaction is presented in 
Fig.~\ref{fig:spp}, which essentially corresponds to the $\Sigma^- n$ case.  
Results are shown for a source radius of $R=1.2$~fm, utilizing 
NLO19(600) (solid line) and NLO(sim) (dash-dotted line), respectively.  
The contributions of the $^1S_0$ and $^3S_1$ partial waves alone (cf. upper 
and lower curves) are indicated by dashed and dotted lines, respectively. 
One can clearly see that there is a strong interplay between the two $S$-wave 
contributions. Moreover, and as expected, there is a sizable difference in 
the overall magnitude of $C(k)$ for the two potentials. 

Actual results for $\Sigma^+ p$, i.e. with the Coulomb interaction included, are 
presented in Fig.~\ref{fig:spp1} for a source radius of $R=1.2$~fm (left) and of
$R=2.5$~fm (right), respectively. Note that the former value corresponds to what
one expects in high-energetic $pp$ collisions as those studied by the ALICE
Collaboration \cite{Acharya:2019,Acharya:2021}
whereas the latter is more characteristic for heavy-ion collisions 
as performed by the STAR Collaboration \cite{Adams:2006}.  
As expected, the Coulomb repulsion between $\Si^+$ and the proton suppresses the 
correlation function at very small momenta and, thus, also the differences
in the $^1S_0$ and the  $^3S_1$ amplitudes. However, for 
$R=1.2$~fm there is definitely a range of momenta, say from $25$ to $75$~MeV/c, 
with a pronounced difference and where pertinent measurements could allow one to
discriminate between the interactions and, thus, facilitate the determination 
of the singlet- and the triplet $S$-wave amplitudes in combination with the 
measured $\Si^+p$ cross section. 
 
There is a visible signal for $R=2.5$~fm, too, see Fig.~\ref{fig:spp1} (right).  
However, it is overall smaller so that data with rather high precision
are needed for distinguishing between the differences in the 
singlet- and triplet amplitudes.  

%%%%%%%%%%%%%%%%%%%%%%%%%%%%%%%%%%%%%%%%%%%%%%%%%%%%%%%%%%%%%%%
{\bf Conclusions:}~ In the present paper we have shown that a determination of 
the spin dependence of two-body interactions like $\Si^+ p$ can be  
achieved by measuring the pertinent momentum correlation function. 
It is feasible under the premises that there is a sign difference 
in the scattering amplitudes of the two spin states involved  
and that the integrated reaction cross section is known. 
The latter is fulfilled for the $\Si^+ p$ system, while for the
former there is strong evidence not least from lattice QCD 
simulations. 

With the $\Si^+ p$ interaction determined, also the one in the $\Si^- n$ 
system is practically fixed, since effects of charge symmetry breaking 
are expected to be small. This would have consequences for the already 
mentioned ongoing discussion on the properties of neutron starts and the 
role played by hyperons for the equation of state.
At present the relevance of the $\Si^-$ is still unsettled \cite{Bombaci:2016}, 
and in some scenarios the $\Si^-$ appears at similar density in neutron stars 
as the $\Lambda$ \cite{Fabbietti:2021NS}. Solid constraints on the strength
of the $\Si^- n$ interaction that can be deduced from measuring and analyzing
the $\Si^+ p$ correlation function would provide here a definite answer. 

Be aware that measurements of the $\Si^+p$ correlation function are not
easily done. But, of course, this is likewise true for conventional scattering
experiments involving the $\Si^+$~\cite{J-PARCE40:2021}. 

Can the idea proposed here be exploited also in studies of other interactions? 
In principle, it should be applicable to any system with only 
two $S$-wave states, i.e. for the scattering of two octet baryons or 
of an octet and a decuplet baryon. 
There are several candidates in the strangeness $S=-2$ to $-4$
sectors where chiral EFT and/or lattice QCD, but also phenomenological
models, predict opposite signs for the $^1S_0$ and $^3S_1$ scattering
lengths. For example, it is the case for $\Xi^0 p$ 
\cite{Fujiwara:2006,Haidenbauer:2018X,HALQCD:2019}, 
for the $S=-3$ system $\Xi^-\La$ \cite{Fujiwara:2006},
and also for $\Xi^0\Xi^-$ \cite{Fujiwara:2006,Doi:2018}. 

\vskip 0.3cm 
%\acknowledgments{
{\it Acknowledments:}
This work is supported in part by the NSFC and the Deutsche Forschungsgemeinschaft (DFG, German Research
Foundation) through the funds provided to the Sino-German Collaborative
Research Center TRR110 ``Symmetries and the Emergence of Structure in QCD''
(NSFC Grant \break No. 12070131001, DFG Project-ID 196253076 - TRR 110). 
The work of UGM was supported in part by the Chinese
Academy of Sciences (CAS) President's International
Fellowship Initiative (PIFI) (Grant No. 2018DM0034,
by VolkswagenStiftung (Grant No. 93562) and by
the European Research Council (ERC) under the European Union's Horizon 2020 research and
innovation programme (EXOTIC, grant agreement No. 101018170).
%}

%%%%%%%%%%%%%%%%%%%%%%%%%%%%%%%%%%%%%%%%%%%%%%%%%%%%%%%%%%%%%%%

\end{document}